\documentclass[10pt,twocolumn,amsmath,amssymb,aps,prl,showpacs,longbibliography,superscriptaddress]{revtex4-1}
\usepackage[latin9]{inputenc}
\usepackage{mathrsfs}
\usepackage{amsmath}
\usepackage{amssymb}
\usepackage{graphicx}
\usepackage{wasysym}
\usepackage{esint}

\makeatletter

\newcommand*\LyXThinSpace{\,\hspace{0pt}}

\usepackage{amsfonts}
\usepackage{graphics}

\makeatother

\begin{document}
\title{Highly anisotropic collective modes of altermagnetic superconductors
with Bogoliubov Fermi surfaces}
\author{Huaisong Zhao}
\affiliation{Centre for Theoretical and Computational Physics, College of Physics,
Qingdao University, Qingdao 266071, P. R. China}
\author{Peng Zou}
\affiliation{Centre for Theoretical and Computational Physics, College of Physics,
Qingdao University, Qingdao 266071, P. R. China}
\author{Xia-Ji Liu}
\affiliation{Centre for Quantum Technology Theory, Swinburne University of Technology,
Melbourne 3122, Australia}
\author{Hui Hu}
\affiliation{Centre for Quantum Technology Theory, Swinburne University of Technology,
Melbourne 3122, Australia}
\date{\today}
\begin{abstract}
We investigate collective modes in charge-neutral altermagnetic superconductors
with $d$-wave spin-split bands, focusing on the impact of emergent
Bogoliubov Fermi surfaces. In the absence of a magnetic field, the
BCS state supports collective modes that inherit the underlying $d$-wave
symmetry, resulting in pronounced momentum-space anisotropy. The broken
time-reversal symmetry further couples these modes to spin-density
fluctuations, allowing their signatures to emerge in the spin dynamical
structure factor. Upon applying a magnetic field, Bogoliubov Fermi
surfaces emerge and qualitatively reshape the collective-mode dynamics
through gapless particle-hole excitations and strongly anisotropic
Landau damping. The damping is strongest along the momentum-space
diagonal, where it can suppress the low-energy phonon mode at small
momenta. Remarkably, the suppressed phonon is accompanied by the emergence
of an in-gap structure in the Higgs mode and an associated pronounced
enhancement of the spin response. These results establish Bogoliubov
Fermi surfaces as a mechanism for generating highly anisotropic collective-mode
dynamics and identify spin response as a sensitive probe of in-gap
Higgs excitations in altermagnetic superconductors.
\end{abstract}
\maketitle
\textbf{\textit{Introduction}}---Altermagnetic superconductivity
has recently emerged as a promising direction for exploring novel
superconducting states arising from the interplay between magnetic
order and Cooper pairing \citep{Mazin2025,Fukaya2025,Liu2026ROPP}.
In contrast to conventional superconductors with spin-degenerate electronic
bands, altermagnetic superconductors exhibit momentum-dependent spin
splitting while retaining zero net magnetization \citep{Hayami2019,Smejkal2020,Smejkal2022,Mazin2023,Amin2024,Jungwirth2025}.
The spin splitting can display nontrivial angular structures, such
as the characteristic $d$-wave form dictated by crystal symmetry,
and therefore intertwines spin, momentum, and superconducting pairing
in a fundamentally different manner from previously established superconductors
\citep{Ouassou2023,Brekke2023,Zhang2024,Lu2024,Chakraborty2024,deCarvalho2024,Hong2025,Sim2025,Sumita2025,Hu2025PRB,Hu2025AB,Liu2026PRB,Liu2026AB,Fu2026,boFu2026}.
This distinctive band structure not only reshapes the quasiparticle
spectrum but also provides a new mechanism for modifying the collective
dynamics of the superconducting condensate.

A particularly striking consequence is the emergence of Bogoliubov
Fermi surfaces (BFSs), extended zero-energy manifolds that provide
an intrinsically gapless phase space for particle-hole excitations
\citep{Hong2025,Hu2025PRB,Liu2026PRB,boFu2026}. In general, BFSs
can emerge in time-reversal-symmetry-breaking superconductors through
the splitting of Bogoliubov bands, inflating an otherwise isolated
point or line node into a two-dimensional zero-energy manifold that
can be protected by a $Z_{2}$ invariant associated with the Pfaffian
of the Bogoliubov-de Gennes Hamiltonian \citep{Agterberg2017,Brydon2018,Dutta2021}.
Their geometry is generically anisotropic, determined by the interplay
between the normal-state dispersion and the momentum dependence of
the superconducting and magnetic order. In altermagnetic superconductors,
the crystal-symmetry-dependent spin splitting provides a distinct
route to BFS formation \citep{Hong2025,Liu2026PRB}, even for conventional
$s$-wave pairing, while preserving zero net magnetization. Their
geometry and anisotropy are then directly inherited from the momentum-space
structure of the altermagnetic order, providing a particularly clean
and general setting for exploring how anisotropic magnetic order affects
superconducting collective dynamics. 

In this Letter, we address the intriguing question of how the gapless
particle-hole excitations of BFSs modify the collective modes in a
charge-neutral altermagnetic superconductor with $d$-wave spin splitting.
In conventional superconductors, the low-energy collective dynamics
are governed by fluctuations of the superconducting order parameter,
most notably the Bogoliubov-Anderson phonon (phase) mode \citep{Anderson1958}
and the Higgs (amplitude) mode \citep{Littlewood1982,Shimano2020}.
For a fully gapped BCS state, the absence of low-energy quasiparticle
excitations suppresses decay into the particle-hole continuum, allowing
these modes to remain well defined at low energies. BFSs change this
situation qualitatively by providing a gapless phase space for coupling
particle-hole excitations to collective fluctuations and transferring
their momentum-space magnetic anisotropy to collective-mode dynamics. 

As a consequence, we find that the presence of BFSs in an altermagnetic
superconductor gives rise to strongly anisotropic Landau damping of
the phonon mode, with the strongest attenuation occurring along the
momentum-space diagonal and leading to a complete suppression of the
low-energy phonon at small momenta. Meanwhile, the gapless quasiparticle
continuum induces a pronounced in-gap structure in the Higgs mode,
accompanied by an enhanced spin-density response that provides a direct
probe of the Higgs excitation. These findings establish a direct link
between the momentum-space structure of the altermagnetic spin splitting,
the geometry of the BFS, and the resulting collective dynamics, and
demonstrate that BFS can give rise to highly anisotropic phonon damping
and an enhanced spin response of the in-gap Higgs mode as distinctive
signatures of its existence in altermagnetic superconductors.

\begin{figure}
\begin{centering}
\includegraphics[width=0.5\textwidth]{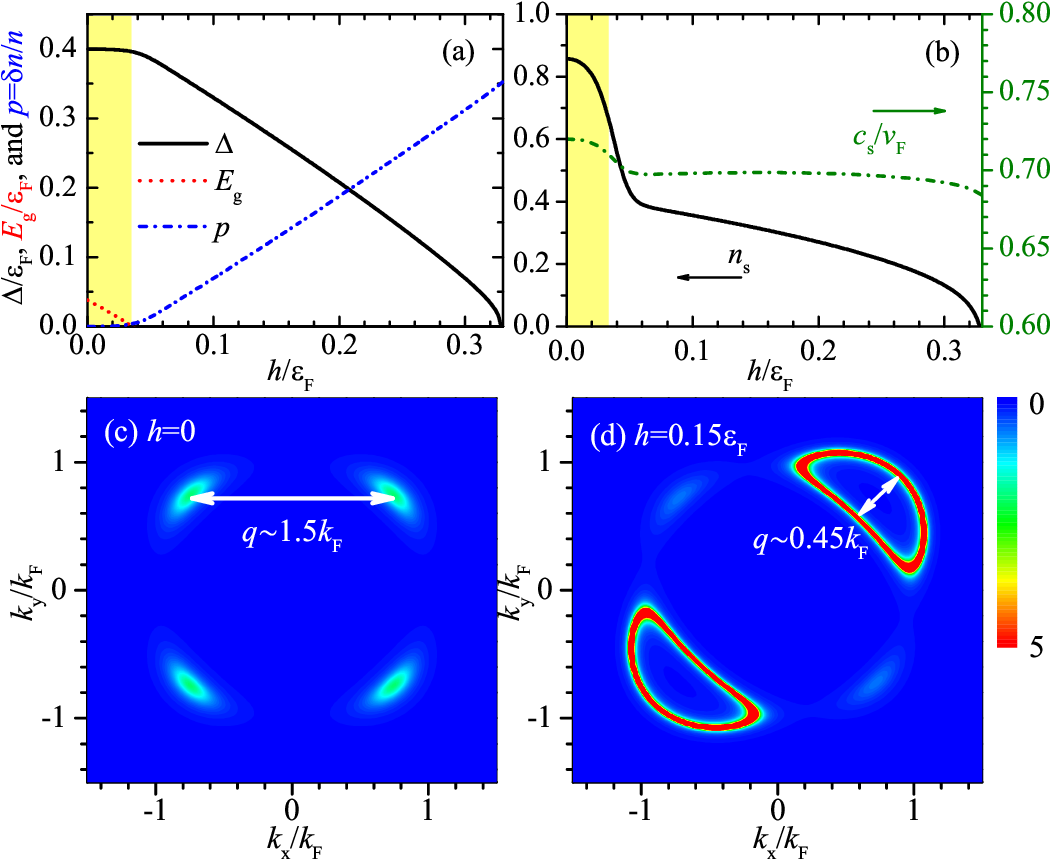}
\par\end{centering}
\caption{\label{fig1: phasediagram} (a) Pairing gap $\Delta$, energy gap
$E_{g}$, spin polarization $\delta n/n$, and (b) superfluid fraction
$n_{s}/n$ and sound velocity $c_{s}$, as functions of the external
magnetic field $h$ for an altermagnetic coupling $\lambda=0.7$.
In each subplot, the yellow shaded region denotes the balanced BCS
state. The subplots (c) and (d) show the spectral function $A(\mathbf{k},\omega=0)$
in the BCS state ($h=0$) and polarized BCS state ($h=0.15\varepsilon_{F}$),
respectively. In the polarized BCS state, the two BFSs are clearly
visible. Throughout the work, we take the binding energy $\varepsilon_{B}=0.08\varepsilon_{F}$
and a negligible temperature $T=0.01T_{F}$.}
\end{figure}

\textbf{\textit{Theoretical framework}}---We start by considering
a minimal single-band model Hamiltonian for a two-dimensional altermagnetic
metal subjected to an external magnetic field $h$ \citep{Hu2025PRB},
$\mathscr{H}=\int d{\bf x}({\cal H}_{0}+{\cal H}_{int})$, where,
with $s_{\uparrow,\downarrow}=\pm1$,
\begin{eqnarray}
{\cal H}_{0} & = & \sum_{\sigma=\uparrow,\downarrow}\psi_{\sigma}^{\dagger}\left({\bf x}\right)\left[\hat{\xi}_{\mathbf{k}}+s_{\sigma}\left(\hat{J}_{\mathbf{k}}+h\right)\right]\psi_{\sigma}\left({\bf x}\right),\\
{\cal H}_{int} & = & U_{0}\psi_{\uparrow}^{\dagger}\left({\bf x}\right)\psi_{\downarrow}^{\dagger}\left({\bf x}\right)\psi_{\downarrow}\left({\bf x}\right)\psi_{\uparrow}\left({\bf x}\right).
\end{eqnarray}
Here, $\hat{\xi}_{\mathbf{k}}\equiv\hbar^{2}\mathbf{\hat{k}}^{2}/(2m)-\mu$
with $\hat{k}_{x,y}=-i\partial_{x,y}$, and $\hat{J}_{\mathbf{k}}\equiv\lambda\hbar^{2}\hat{k}_{x}\hat{k}_{y}/(2m)$
represents the altermagnetic term with a $d$-wave form and $C_{4z}$
rotational symmetry. The chemical potential $\mu$ is determined self-consistently
by the total particle density $n=n_{\uparrow}+n_{\downarrow}$. Superconductivity
is induced by the attractive interparticle interaction of strength
$U_{0}$, which is regularized in terms of the two-body binding energy
$\varepsilon_{B}$ according to $U_{0}^{-1}=-(1/V)\sum_{\mathbf{k}}[\hbar^{2}k^{2}/m+\varepsilon_{B}]^{-1}$,
where $V$ denotes the area of the system \citep{He2015}. In this
work, we focus on a charge-neutral Fermi system, motivated by a possible
realization of the model using ultracold atoms \citep{Das2024}. Alternatively,
the model might be realized in an excitonic system by engineering
altermagnietsm \citep{Cao2025,Wang2026}, thereby providing a route
towards an altermagnetic excitonic insulator.

The interplay between the altermagnetic coupling and the external
magnetic field leads to a rich phase diagram, as discussed in previous
studies \citep{Chakraborty2024,Hu2025PRB}. In Fig. \ref{fig1: phasediagram},
we show the evolution from a fully gapped BCS phase to a gapless polarized
BCS state with increasing magnetic field for an altermagnetic coupling
$\lambda=0.7$. The polarized BCS state is characterized by a finite
spin-population imbalance, $\delta n=n_{\uparrow}-n_{\downarrow}\neq0$,
accompanied by the closing of the energy gap, $E_{g}=0$. At the same
time, as shown in Fig. \ref{fig1: phasediagram}(d), two BFSs emerge
along the diagonal directions of momentum space, as revealed by the
zero-energy spectral function $A(\mathbf{k},\omega=0)=-\textrm{Im\ensuremath{(\mathcal{G}_{11}^{(0)}+\mathcal{G}_{22}^{(0)})/\pi}}$,
obtained after the analytic continuation $i\omega_{m}\rightarrow\omega+i0^{+}$.
Here, $\mathcal{G}_{ij}^{(0)}$ ($i,j=1,2$) denotes the matrix elements
of the $2\times2$ BCS Green's function $\mathcal{G}_{0}(k)$, with
four-momentum $k=(\mathbf{k},i\omega_{m})$,
\begin{equation}
\mathcal{G}_{0}=\left[\begin{array}{cc}
i\omega_{m}-\xi_{\mathbf{k}}-J_{\mathbf{k}}-h & -\Delta\\
-\Delta & i\omega_{m}+\xi_{\mathbf{k}}-J_{\mathbf{k}}-h
\end{array}\right]^{-1}.\label{eq:gfBCS}
\end{equation}
At the representative magnetic field $h=0.15\varepsilon_{F}$, the
two BFSs exhibit a pronounced anisotropy, extending over a width of
approximately $q\sim0.5k_{F}$ along the diagonal direction. Their
appearance gives rise to substantial changes in the superfluid density
$n_{s}$ along the $x$-axis and the adiabatic sound velocity $c_{s}$,
as illustrated in Fig. \ref{fig1: phasediagram}(b). These changes
are also reflected in the collective dynamical responses, which we
discuss below.

To describe the collective dynamics, we employ a functional path-integral
formalism within the random phase approximation (RPA) \citep{He2016}.
We introduce the source fields $j_{c}(\mathbf{x},\tau)$ and $j_{s}(\mathbf{x},\tau)$,
which couple to the charge-density and spin-density, respectively.
The quartic four-fermion interaction is decoupled via an auxiliary
Hubbard-Stratonvich pairing field, after which the fermionic degrees
of freedom are integrated out to obtain an effective action $\mathcal{S}_{\textrm{eff}}[j_{c},j_{s}]$
for the bosonic pairing fluctuations \citep{He2015,Ohashi2002,Hu2006,Diener2008}.
Truncating the effective action at the Gaussian level, corresponding
to the RPA, and taking its second-order functional derivative with
respect to the source fields \citep{He2016}, we obtain the density
response functions $\chi_{\nu\nu}^{\textrm{RPA}}(\mathbf{q},\omega)$,
where $\nu=c,s$ labels the charge and spin channels, respectively:\begin{widetext}
\begin{equation}
\chi_{\nu\nu}^{\textrm{RPA}}\left(\mathbf{q},\omega\right)=\chi_{\nu\nu}\left(\mathbf{q},\omega\right)+\left[\begin{array}{cc}
\chi_{1\nu}\left(\mathbf{q},\omega\right), & \chi_{1\nu}^{*}\left(-\mathbf{q},-\omega\right)\end{array}\right]\left[\begin{array}{cc}
\Gamma_{11}\left(\mathbf{q},\omega\right) & \Gamma_{12}\left(\mathbf{q},\omega\right)\\
\Gamma_{12}\left(\mathbf{q},\omega\right) & \Gamma_{22}\left(\mathbf{q},\omega\right)
\end{array}\right]\left[\begin{array}{c}
\chi_{1\nu}\left(\mathbf{q},\omega\right)\\
\chi_{1\nu}^{*}\left(-\mathbf{q},-\omega\right)
\end{array}\right].\label{eq:kappaRPA}
\end{equation}
\end{widetext}Here, $\chi_{\nu\nu}(\mathbf{q},\omega)$ denotes the
bare particle-hole bubble diagram evaluated using the BCS Green's
function $\mathcal{G}_{0}$. The second term accounts for the repeated
particle-hole scattering processes to all orders, thereby incorporating
the dynamical feedback of collective pair fluctuations required to
preserve gauge invariance in the superconducting response \citep{Nambu1960}.
These collective effects are encoded in the $2\times2$ pairing vertex
function $\boldsymbol{\Gamma}$, which can be viewed as the propagator
of Cooper pairs \citep{Ohashi2002,Hu2006,Diener2008}. The explicit
expressions for $\chi_{\nu\nu}(\mathbf{q},\omega)$, $\chi_{1\nu}(\mathbf{q},\omega)$
($\nu=c,s$), and $\Gamma_{ij}(\mathbf{q},\omega)$ ($i,j=1,2$) are
provided in the Appendix. To identify the collective modes, we resolve
$\boldsymbol{\Gamma}$ into its phase and amplitude components \citep{Ohashi2002,Diener2008},
$\Gamma_{\theta\theta}=(\Gamma_{11}+\Gamma_{22})/2-\Gamma_{12}$ and
$\Gamma_{\lambda\lambda}=(\Gamma_{11}+\Gamma_{22})/2+\Gamma_{12}$.
The phonon and Higgs modes are then characterized by the corresponding
unnormalized spectral functions $A_{\theta\theta}=-\textrm{Im}\Gamma_{\theta\theta}/\pi$
and $A_{\lambda\lambda}=-\textrm{Im}\Gamma_{\lambda\lambda}/\pi$,
respectively.

\begin{figure}
\begin{centering}
\includegraphics[width=0.25\textwidth]{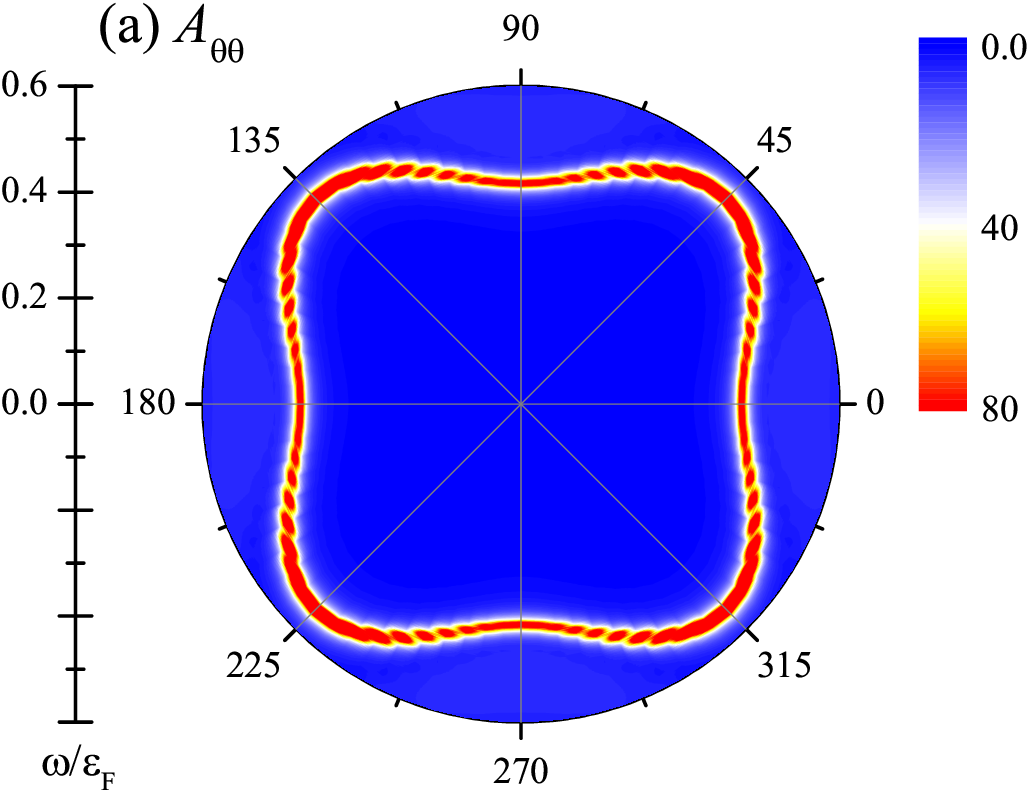}\includegraphics[width=0.25\textwidth]{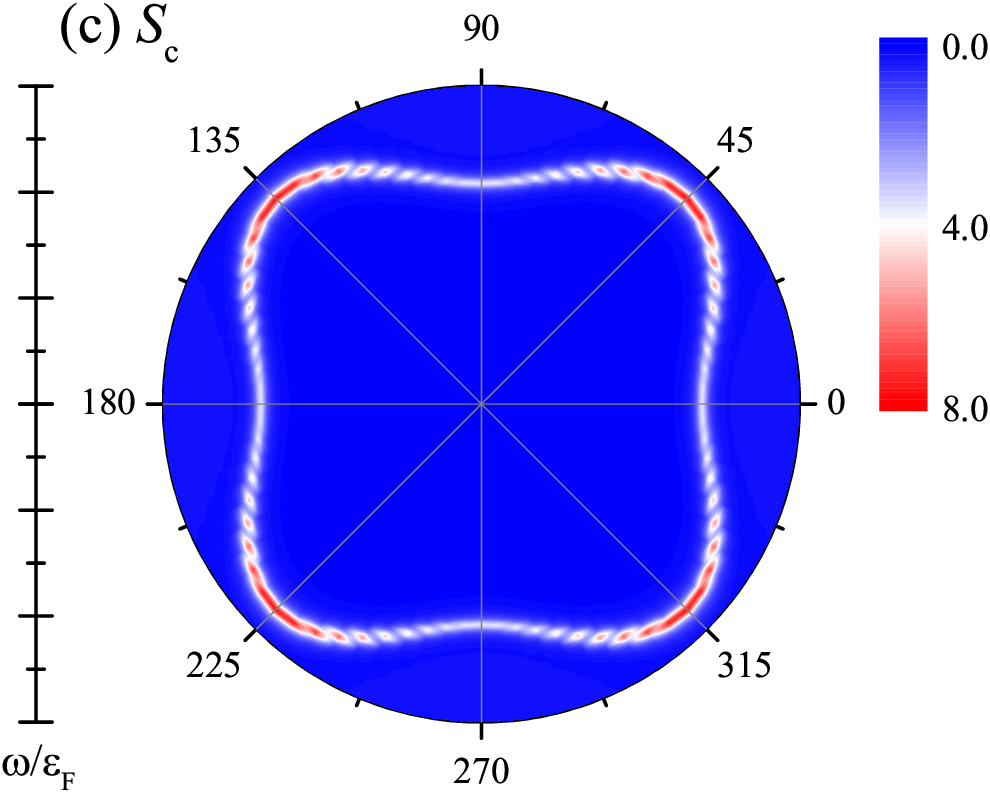}
\par\end{centering}
\begin{centering}
\includegraphics[width=0.25\textwidth]{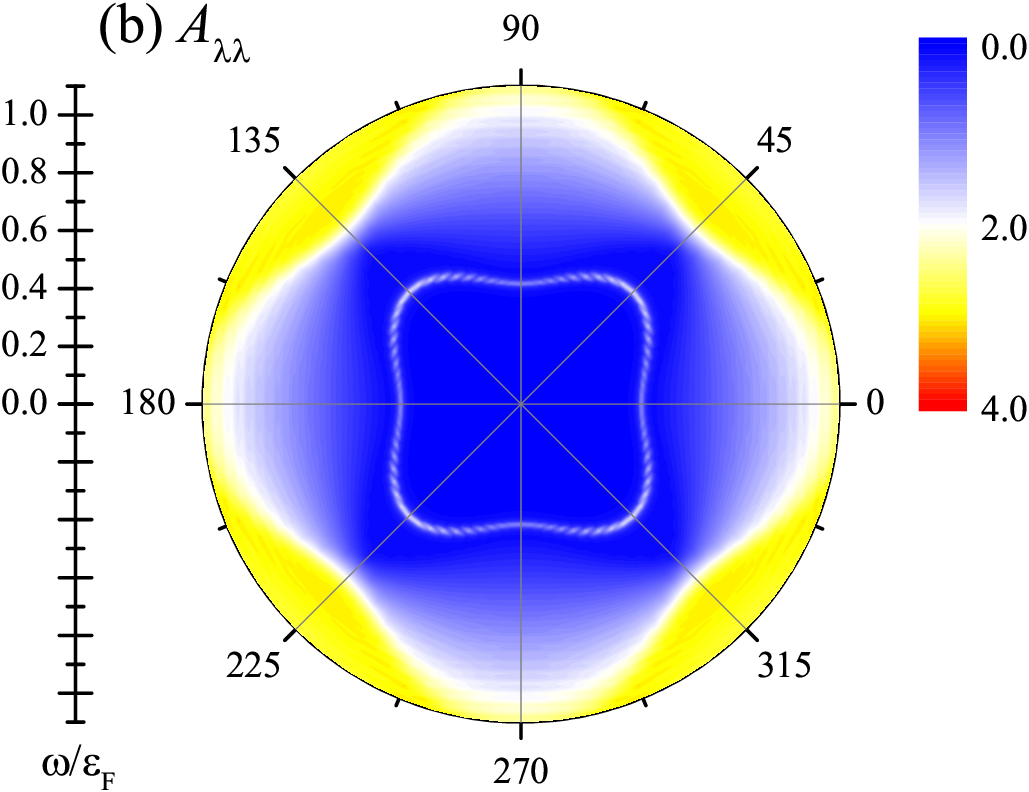}\includegraphics[width=0.25\textwidth]{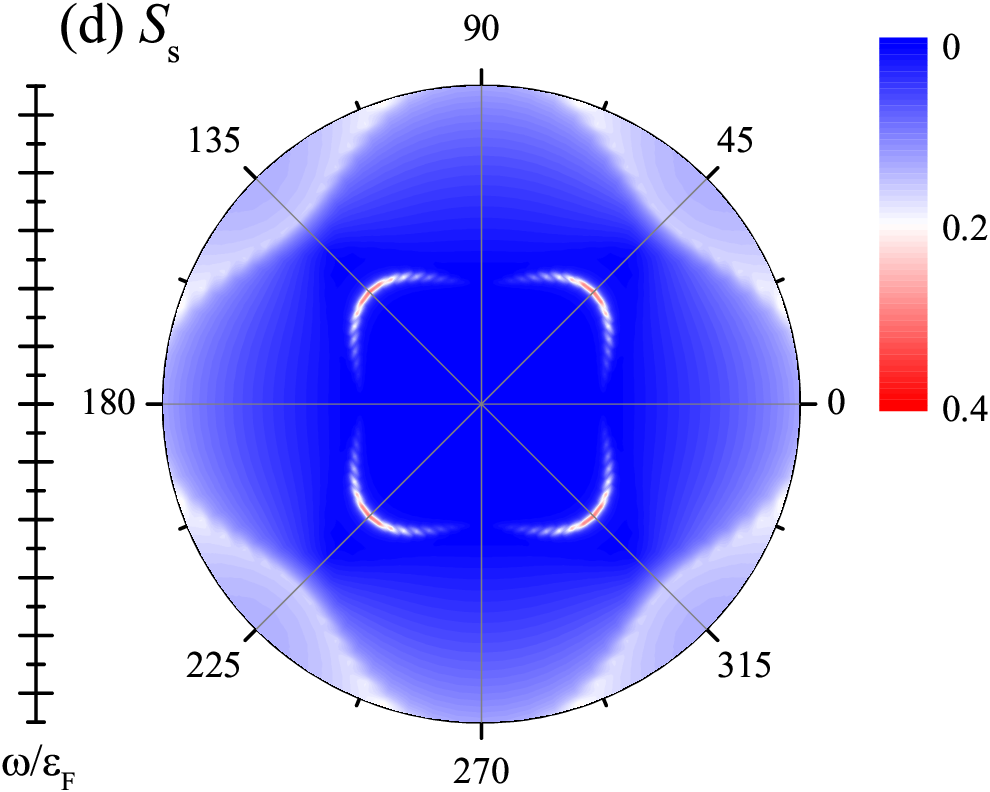}
\par\end{centering}
\caption{\label{fig2: BCS} Polar contour plots of the Cooper pair spectral
functions $A_{\theta\theta}(\mathbf{q},\omega)$ (a) and $A_{\lambda\lambda}(\mathbf{q},\omega)$
(b), together with the charge DSF $S_{c}(\mathbf{q},\omega)$ (c)
and the spin DSF $S_{s}(\mathbf{q},\omega)$ (d), for the fully gapped
BCS phase with $\lambda=0.7$ and $h=0$. In each panel, the radial
coordinate represents $\omega/\varepsilon_{F}$, as indicated on the
left, while the angular coordinate gives the direction of the momentum
$\mathbf{q}$, $\varphi_{\mathbf{q}}$, in radians. The spectral-function
colormaps are shown in arbitrary units, whereas the DSF colormaps
are expressed in units of $\varepsilon_{F}/n$. The momentum magnitude
is fixed at $q=\left|\mathbf{q}\right|=0.5k_{F}$. Throughout the
work, a small spectral broadening of $0.01\varepsilon_{F}$ is introduced
to render the $\delta$-function peaks visible in the plots.}
\end{figure}

\textbf{\textit{Fully gapped BCS phases}}---We first consider the
collective modes of the fully gapped BCS phase with an altermagnetic
coupling $\lambda=0.7$ and in the absence of external magnetic field.
Figure \ref{fig2: BCS} reports the pair spectral function $A_{\theta\theta}(\mathbf{q},\omega)$
and $A_{\lambda\lambda}(\mathbf{q},\omega)$ at $\left|\mathbf{q}\right|=0.5k_{F}$,
together with the charge and spin dynamic structure factors (DSFs),
defined as $S_{c,s}(\mathbf{q},\omega)=-[1+n_{B}(\omega)]\textrm{Im}\chi_{cc,ss}^{\textrm{RPA}}(\mathbf{q},\omega)/\pi$,
where $n_{B}(\omega)\equiv1/[e^{\hbar\omega/(k_{B}T)}-1]$ is the
Bose-Einstein distribution function. The results are presented as
polar contour plots, with the radial coordinate representing the frequency
$\omega$ and the angular coordinate specifying the direction of the
momentum $\mathbf{q}$, denoted by $\varphi_{\mathbf{q}}$.

An immediate observation is that the spectral functions of both the
phonon and Higgs modes inherit the $C_{4z}$ rotational symmetry of
the underlying $d$-wave altermagnetism. Accordingly, the phonon dispersion
is strongly direction dependent: its frequency is minimized along
the $x$- and $y$-axes and maximized along the diagonal directions.
As the phonon couples directly to charge-density fluctuations \citep{Anderson1958},
this anisotropy is directly imprinted on the charge DSF $S_{c}(\mathbf{q},\omega)$,
as shown in Fig. \ref{fig2: BCS}(c).

More strikingly, the spin DSF exhibits a qualitatively distinct response
in the presence of $d$-wave altermagnetism {[}Fig. \ref{fig2: BCS}(d){]}.
In conventional superconductors, the spin response primarily probes
quasiparticle excitations, such as the pair-breaking continuum, while
the collective phonon and Higgs modes are decoupled from the spin
channel \citep{Anderson1958,Littlewood1982,Shimano2020}. This decoupling
follows from the vanishing mixed susceptibility $\chi_{1s}(\mathbf{q},\omega)=0$
imposed by time-reversal symmetry (see Appendix for more discussions).
Here, however, the time-reversal-symmetry-breaking altermagnetic order
renders $\chi_{1s}(\mathbf{q},\omega)$ finite, thereby enabling the
collective modes - most notably the phonon mode, as discussed in the
Appendix - to acquire a direct signature in the spin DSF within the
pair-breaking gap.

\begin{figure}
\begin{centering}
\includegraphics[width=0.5\textwidth]{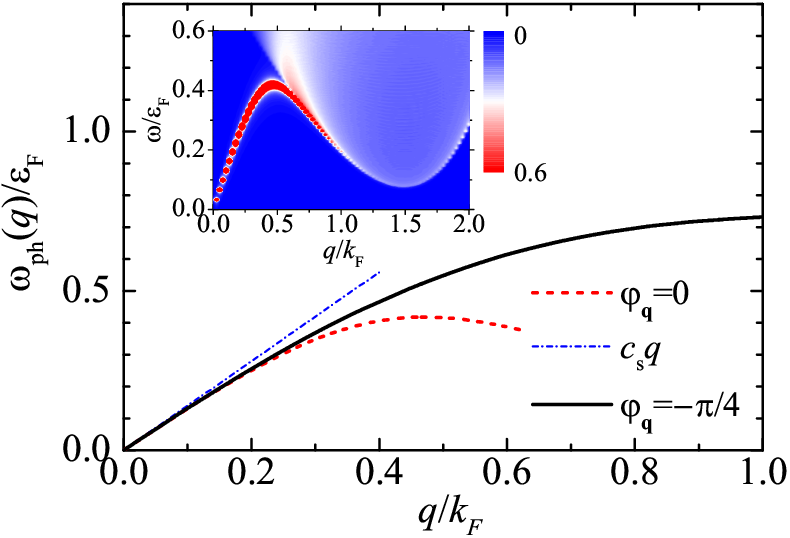}
\par\end{centering}
\caption{\label{fig3: w0BCS} Phonon dispersion relations in the fully gapped
BCS phase with $\lambda=0.7$ and $h=0$ for two momentum directions,
$\varphi_{\mathbf{q}}=-45^{\circ}$ (black solid line) and $\varphi_{\mathbf{q}}=0$
(red dashed line), compared with the low-momentum linear dispersion
$c_{s}q$ (blue dot-dashed line). The inset shows the contour plot
of the charge DSF, $S_{c}(\mathbf{q},\omega)$, at $\varphi_{\mathbf{q}}=-45^{\circ}$.
Despite the anisotropy of the dispersion, the microscopic sound velocity,
defined by $\lim_{q\rightarrow0}[\omega_{\textrm{ph}}(q)/q]$, is
independent of the propagation direction $\varphi_{\mathbf{q}}$ and
coincides with the thermodynamic sound velocity $c_{s}$.}
\end{figure}

In Fig. \ref{fig3: w0BCS}, we show the phonon dispersion $\omega_{\textrm{ph}}(q=\left|\mathbf{q}\right|)$
along two representative momentum directions. While $\omega_{\textrm{ph}}(q)$
increases monotonically along the diagonals, it bends downward at
$q\sim0.5k_{F}$ along the $x$ and $y$ axes, signaling a pronounced
softening. This softening is particularly evident in the charge DSF
(inset), which develops a roton-like minimum at $q\sim1.5k_{F}$.
This roton structure originates from inter-minima particle-hole excitations
{[}Fig. \ref{fig1: phasediagram}(c){]} and signals the instability
toward a spatially modulated Fulde-Ferrell-Larkin-Ovchinnikov superconducting
state at sufficiently strong altermagnetic coupling \citep{Hong2025,Sumita2025,Hu2025PRB,Liu2026PRB},
even in the absence of an external magnetic field.

\begin{figure}
\begin{centering}
\includegraphics[width=0.25\textwidth]{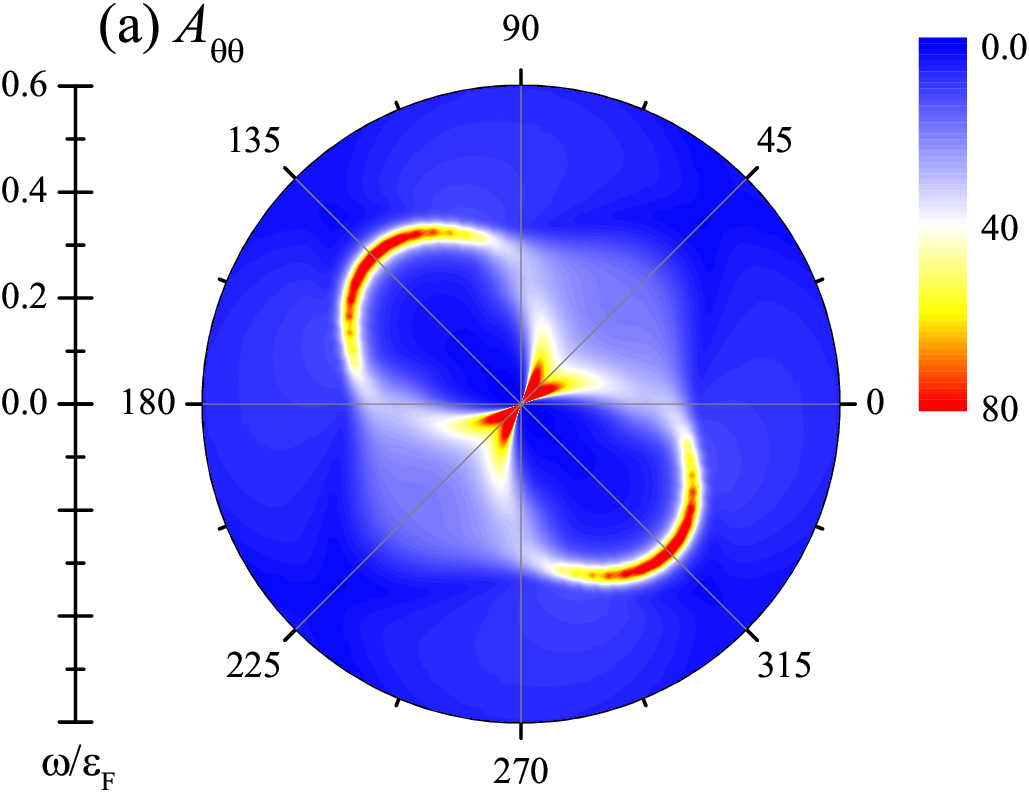}\includegraphics[width=0.25\textwidth]{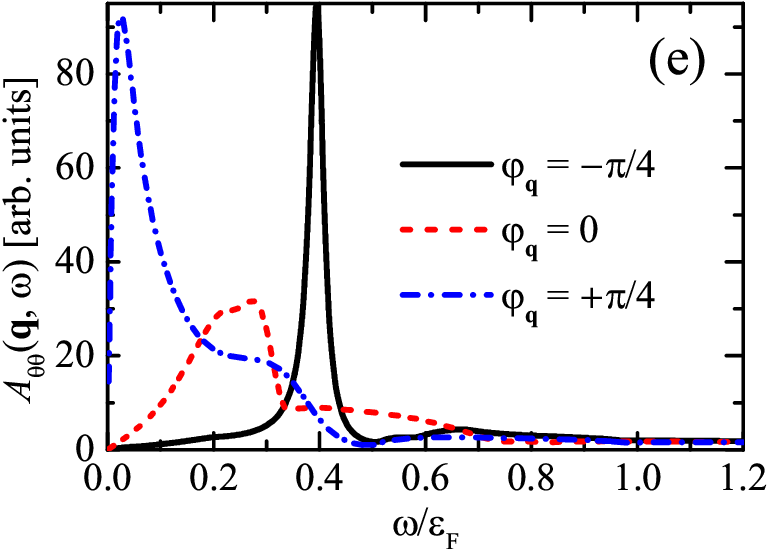}
\par\end{centering}
\begin{centering}
\includegraphics[width=0.25\textwidth]{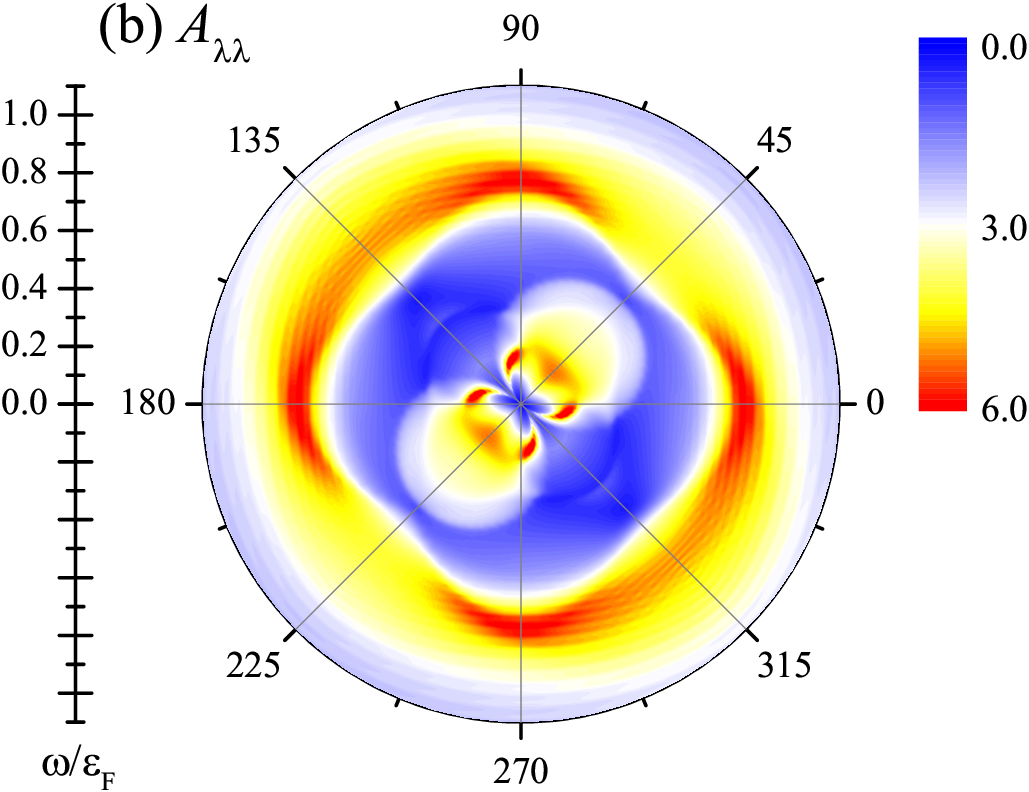}\includegraphics[width=0.25\textwidth]{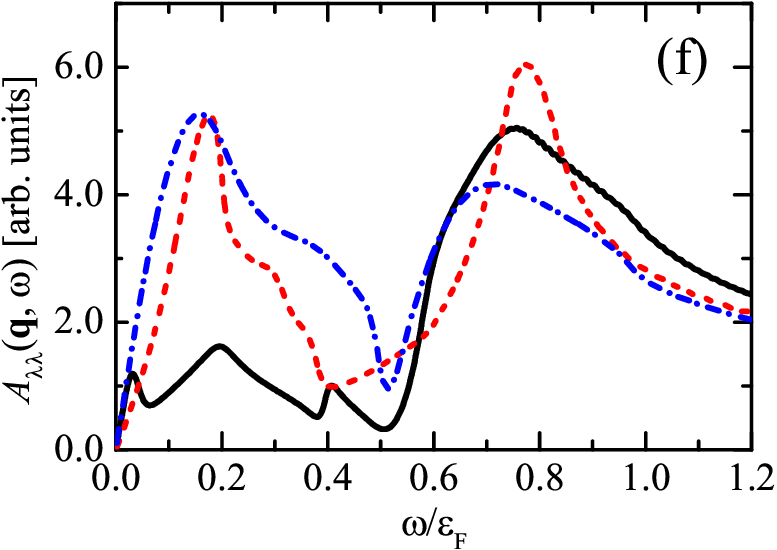}
\par\end{centering}
\begin{centering}
\includegraphics[width=0.25\textwidth]{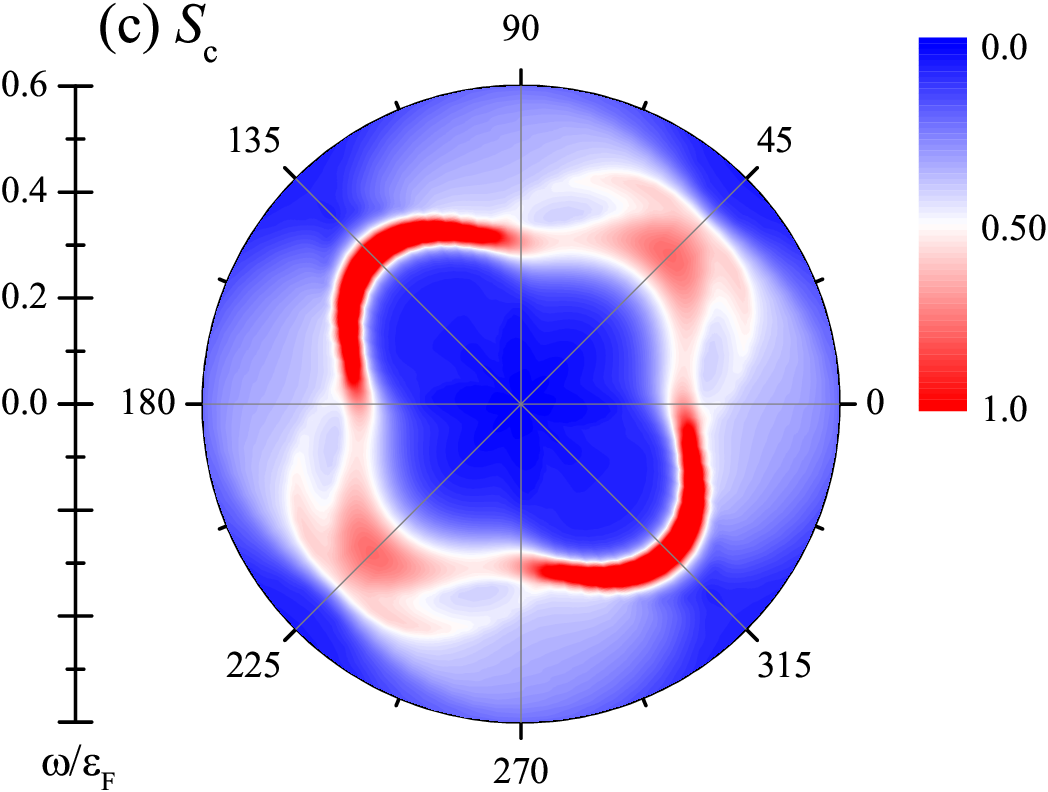}\includegraphics[width=0.25\textwidth]{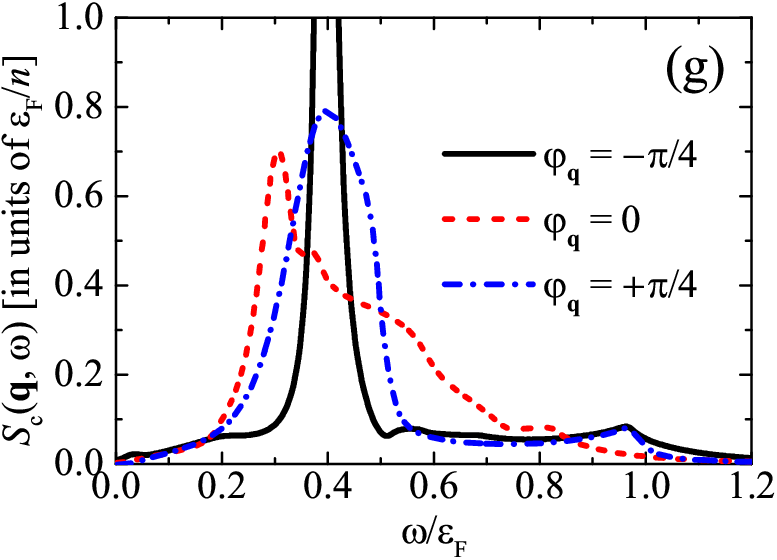}
\par\end{centering}
\begin{centering}
\includegraphics[width=0.25\textwidth]{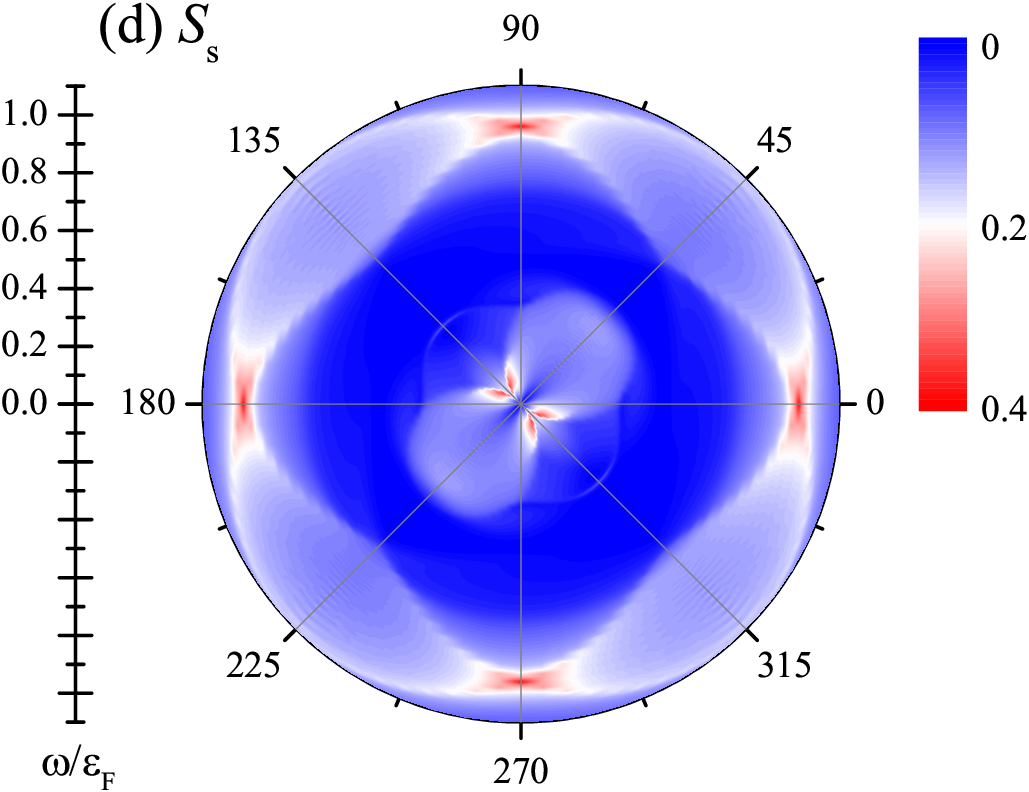}\includegraphics[width=0.25\textwidth]{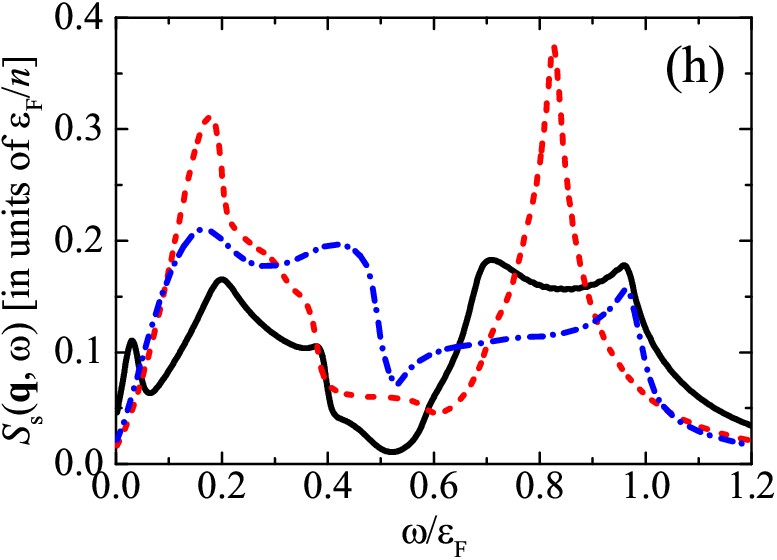}
\par\end{centering}
\caption{\label{fig4: pBCS} Left column: Polar contour plots of $A_{\theta\theta}(\mathbf{q},\omega)$
(a), $A_{\lambda\lambda}(\mathbf{q},\omega)$ (b), $S_{c}(\mathbf{q},\omega)$
(c) and $S_{s}(\mathbf{q},\omega)$ (d) for a polarized BCS state
with $\lambda=0.7$ and $h=0.15\varepsilon_{F}$. Right column: Corresponding
plots for three distinct angular directions of the momentum $\mathbf{q}$,
as indicated in panels (e)-(g). The momentum magnitude is fixed at
$q=\left|\mathbf{q}\right|=0.3k_{F}$. }
\end{figure}

\textbf{\textit{Gapless polarized BCS states}}---Approaching the
gapless polarized BCS regime, the emergence of BFSs strongly reshapes
the collective dynamics. In Fig. \ref{fig4: pBCS}, the left column
shows polar $\omega$-$\varphi_{\mathbf{q}}$ maps of the pair spectral
functions and the charge and spin DSFs for $\lambda=0.7$ and $h=0.15\varepsilon_{F}$
at $\left|\mathbf{q}\right|=0.3k_{F}$, with cuts along representative
directions shown in the right column.

\begin{figure}
\begin{centering}
\includegraphics[width=0.5\textwidth]{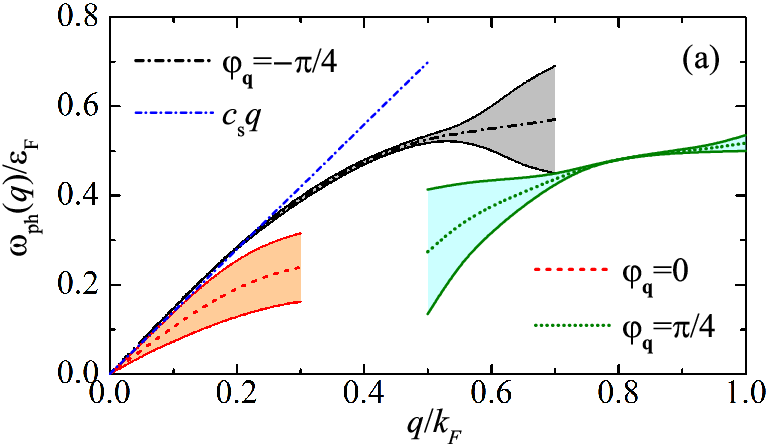}
\par\end{centering}
\begin{centering}
\includegraphics[width=0.25\textwidth]{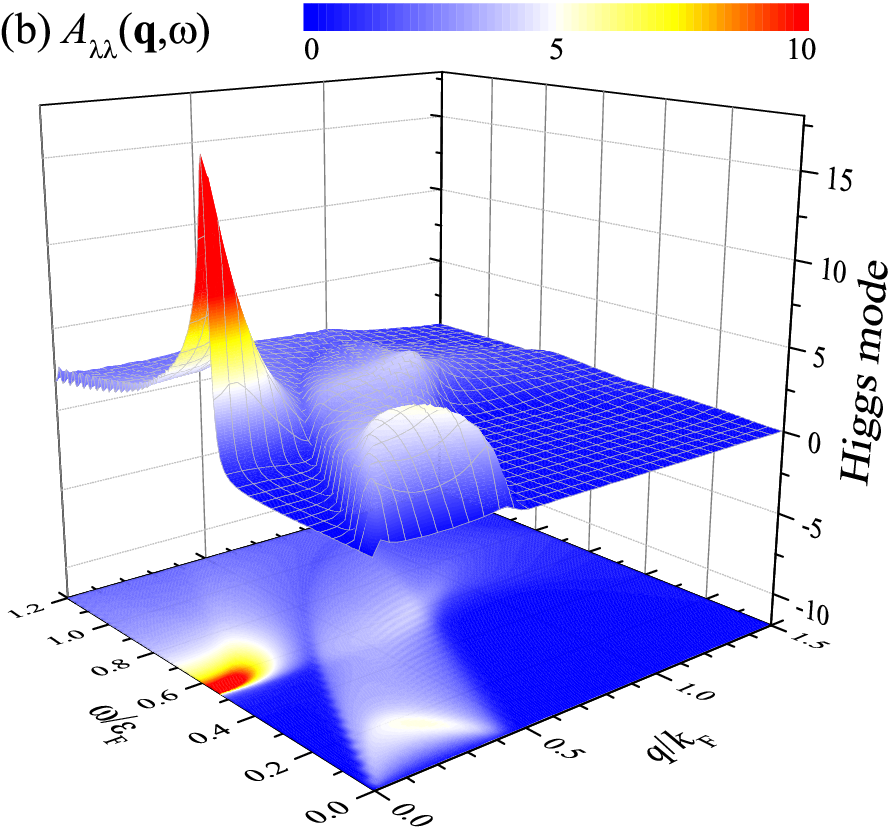}\includegraphics[width=0.255\textwidth]{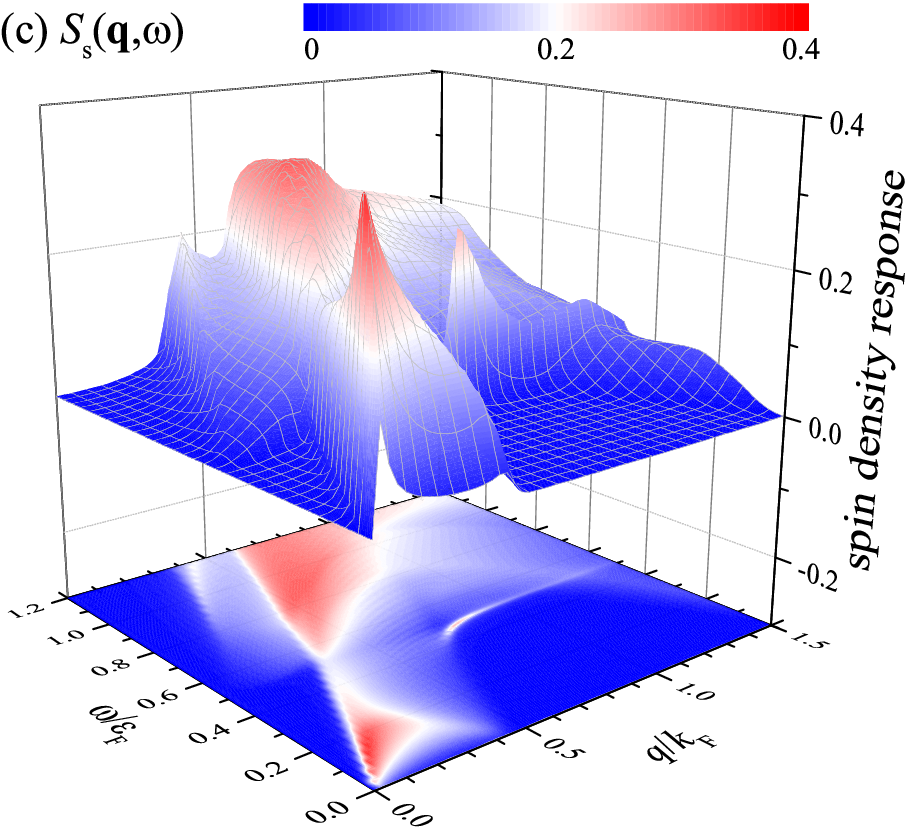}
\par\end{centering}
\caption{\label{fig5: w0pBCS} (a) Phonon dispersion relations in the polarized
BCS state with $\lambda=0.7$ and $h=0.15\varepsilon_{F}$ for three
momentum directions: $\varphi_{\mathbf{q}}=-45^{\circ}$ (black dot-dashed
line), $\varphi_{\mathbf{q}}=0$ (red dashed line) and $\varphi_{\mathbf{q}}=+45^{\circ}$
(green dashed line). The shaded bands denote the peak widths for each
momentum direction, extracted by fitting the phonon spectral function
to a Lorentzian lineshape. (b), (c) Contour plots of the pair spectral
function $A_{\lambda\lambda}(\mathbf{q},\omega)$ and the spin DSF
$S_{s}(\mathbf{q},\omega)$, respectively, along the diagonal momentum
direction, $\varphi_{\mathbf{q}}=+45^{\circ}$.}
\end{figure}

Along the diagonal direction, $\varphi_{\mathbf{q}}=+45^{\circ}$,
both the phonon and Higgs modes are strongly modified. The phonon
mode is almost completely suppressed, acquiring a diffusive character
reminiscent of the dynamics above the superconducting transition,
while the Higgs mode develops a pronounced in-gap feature. These effects
arise from gapless particle-hole excitations, which are particularly
favorable along this direction {[}see Fig. \ref{fig1: phasediagram}(d){]}
for $0<|\mathbf{q}|\lesssim0.45k_{F}$. Particle-hole excitations
near the BFSs also affect both collective modes along the anti-diagonal
direction, $\varphi_{\mathbf{q}}=-45^{\circ}$. However, the strongly
anisotropic BFS geometry substantially restricts the available phase
space in this direction, resulting in only weak damping for $|\mathbf{q}|\apprle0.6k_{F}$.
Thus, the anisotropy of the BFSs is directly reflected in a pronounced
momentum-space anisotropy of the collective dynamics.

In experiments, this pronounced momentum-space anisotropy should be
directly manifested in the charge and spin responses, as shown in
Figs. \ref{fig4: pBCS}(c) and \ref{fig4: pBCS}(d). Along the diagonal
direction, the charge DSF becomes strongly broadened and loses the
well-defined dispersive feature associated with the phonon, providing
a direct signature of its strong Landau damping by gapless particle-hole
excitations. In contrast, the spin DSF develops a pronounced in-gap
structure whose spectral weight closely follows that of the Higgs
mode. As demonstrated in the Appendix, the low-energy peak in the
spin DSF originates from Higgs excitations. The resulting spin-Higgs
coupling therefore provides a direct channel through which the spin
DSF can probe the otherwise weakly visible Higgs sector \citep{Shimano2020}.

To quantify the momentum-space anisotropy of the collective dynamics,
Fig. \ref{fig5: w0pBCS}(a) shows the phonon dispersion and damping
rate up to $|\mathbf{q}|=k_{F}$ for several momentum directions.
Along the off-diagonal direction, $\varphi_{\mathbf{q}}=-45^{\circ}$,
the phonon remains essentially undamped up to $|\mathbf{q}|\simeq0.5k_{F}$,
reflecting the strongly restricted phase space for particle-hole excitations
near the highly anisotropic BFSs discussed above. The situation is
markedly different along the diagonal direction, $\varphi_{\mathbf{q}}=+45^{\circ}$,
where gapless particle-hole excitations strongly disrupt the phonon
at low momentum, such that a well-defined mode emerges only for $|\mathbf{q}|\gtrsim0.5k_{F}$.
Interestingly, the damping becomes negligible again near $|\mathbf{q}|\simeq0.8k_{F}$,
indicating that the phase space available for decay is itself strongly
momentum dependent. The resulting nonmonotonic damping thus provides
a direct fingerprint of the anisotropic BFS geometry and its control
over collective-mode decay. 

Finally, motivated by the close correspondence between the in-gap
spin response and the Higgs spectral function, we directly compare
the two in Figs. \ref{fig5: w0pBCS}(b) and \ref{fig5: w0pBCS}(c)
for $\varphi_{\mathbf{q}}=+45^{\circ}$. The two-dimensional spectra
reveal a close similarity at low momentum, with the in-gap structure
of the spin DSF roughly tracking the Higgs spectral weight. This agreement
further supports the view that the spin response can serve as an alternative
signature of the Higgs mode in the gapless polarized BCS state, despite
the strong Landau damping and anisotropy induced by the BFSs.

\textbf{\textit{Conclusions and outlooks}}---In summary, we have
investigated the collective dynamics of $d$-wave altermagnetic superconductors
across both fully gapped BCS and gapless polarized BCS regimes. In
the fully gapped phase, broken time-reversal symmetry gives rise to
a pronounced spin-density response, providing a distinctive feature
of the collective dynamics. In the gapless polarized BCS regime, the
emergence of highly anisotropic BFSs profoundly reshapes the collective
modes, resulting in strongly momentum-direction-dependent phonon and
Higgs dynamics and, consequently, highly anisotropic charge and spin
responses. We argue that this pronounced anisotropy is a generic consequence
of the momentum-space geometry of the BFSs: the direction-dependent
phase space for gapless particle-hole excitations governs both the
dispersion and damping of the collective modes. The resulting anisotropic
collective dynamics therefore provides a sensitive spectroscopic probe
of the underlying BFSs and their momentum-space structure, beyond
conventional single-particle probes such as angle-resolved photoemission
spectroscopy.

An interesting direction for future work is to incorporate long-range
Coulomb interactions and explore their interplay with the collective
dynamics. In a charged superconductor, the phase mode is converted
into a plasmon through the Anderson-Higgs mechanism \citep{Anderson1958};
however, in the gapless polarized BCS regime, the coexistence of condensate
and gapless quasiparticles may also support a low-energy Carlson-Goldman
mode, even at zero temperature \citep{Carlson1975,Ohashi1997}. Given
the strongly anisotropic Bogoliubov Fermi surfaces, both the dispersion
and damping of this mode are expected to inherit a pronounced momentum-space
anisotropy through the direction-dependent quasiparticle response.
It would therefore be interesting to determine how Coulomb screening
redistributes spectral weight between the plasmon and Carlson-Goldman
sectors, and whether their anisotropic collective dynamics can provide
a complementary probe of the underlying Bogoliubov Fermi-surface geometry.
\begin{acknowledgments}
\textit{Acknowledgments}---We thank Professor Yoji Ohashi for discussions
on the Carlson-Goldman mode. This research was supported by the National
Natural Science Foundation of China under Grant No. U23A2073 (P.Z.).

\textit{Data availability}---The data that support the findings of
the work are openly available \citep{DataLink}.
\end{acknowledgments}

\appendix

\section{RPA response functions}

In this section, we provide a brief derivation of the RPA response
functions, Eq. (\ref{eq:kappaRPA}). As discussed in the main text,
using the standard functional path-integral formalism in imaginary
time \citep{He2015,He2016,Diener2008}, we obtain an effective action
${\cal S}_{\textrm{eff}}[\Delta^{*},\Delta,j]$ for the Cooper pairing
field $\Delta(\mathbf{x},\tau)$ in the presence of an external source
field $j(\mathbf{x},\tau)$. To illustrate the procedure, we take
the source field $j_{c}(\mathbf{x},\tau)$ as an example, which couples
to the total density operator and generates the charge-density response:
\begin{equation}
{\cal S}_{\textrm{eff}}=-\text{Tr}\ln\left[-{\cal G}_{j}^{-1}\right]-\int d\tau d{\bf x}\frac{\Delta^{*}\Delta}{U_{0}}+\beta\sum_{{\bf k}}\hat{\xi}_{{\bf k}}.\label{eq:Seff}
\end{equation}
Here $\beta=1/(k_{B}T)$ is the inverse temperature, and ${\cal G}_{j}({\bf x},\tau;{\bf x}^{\prime},\tau^{\prime})$
is the single-particle Green's function in the presence of both the
fluctuating pairing field and the source field. It can be expressed
in the form,\begin{widetext} 

\begin{equation}
{\cal G}_{j}^{-1}\left({\bf x},\tau;{\bf x}^{\prime},\tau^{\prime}\right)\equiv\left[\begin{array}{cc}
-\partial_{\tau}-\hat{\xi}_{{\bf k}}-\frac{\lambda\hbar^{2}}{2m}\hat{k}_{x}\hat{k}_{y}-h-j_{c}\left({\bf x},\tau\right) & -\Delta\left({\bf x},\tau\right)\\
-\Delta^{*}\left({\bf x},\tau\right) & -\partial_{\tau}+\hat{\xi}_{{\bf k}}-\frac{\lambda\hbar^{2}}{2m}\hat{k}_{x}\hat{k}_{y}-h+j_{c}\left({\bf x},\tau\right)
\end{array}\right]\delta\left({\bf x-x}^{\prime}\right)\delta\left(\tau-\tau^{\prime}\right).
\end{equation}
The trace in Eq. (\ref{eq:Seff}) is taken over all relevant degrees
of freedom, including the Nambu spinor, spatial coordinate (${\bf x}$),
and imaginary time ($\tau$). To proceed, we note that, in the superconducting
phase, the pairing field acquires a nonzero saddle-point value, $\Delta\neq0$.
Therefore, we write $\Delta\left({\bf x},\tau\right)=\Delta+\varphi\left({\bf x},\tau\right)$,
where the saddle-point pairing gap $\Delta$ is chosen to be real,
and decompose the inverse Green's function into the standard BCS contribution
($\mathcal{G}_{0}^{-1}$) and a residual contribution $(-\delta\mathcal{G}^{-1})$
that contains both the pair fluctuations and the source field. Thus,
$\mathcal{G}_{j}^{-1}=\mathcal{G}_{0}^{-1}(1-\mathcal{G}_{0}\delta\mathcal{G}{}^{-1}$),
where 
\begin{equation}
\mathcal{G}_{0}\delta\mathcal{G}^{-1}=\mathcal{G}_{0}\left[\begin{array}{cc}
j_{c}\left({\bf x},\tau\right) & \varphi\left({\bf x},\tau\right)\\
\varphi^{*}\left({\bf x},\tau\right) & -j_{c}\left({\bf x},\tau\right)
\end{array}\right]=\left(\mathcal{G}_{0}\sigma^{-}\right)\varphi^{*}\left({\bf x},\tau\right)+\left(\mathcal{G}_{0}\sigma^{+}\right)\varphi\left({\bf x},\tau\right)+\left(\mathcal{G}_{0}\sigma_{z}\right)j_{c}\left({\bf x},\tau\right),
\end{equation}
and $\sigma^{+,-}=(\sigma_{x}\pm i\sigma_{y})/2$ and $\sigma_{z}$
are the Pauli matrices. In momentum space, the standard BCS Green's
function $\mathcal{G}_{0}(\mathbf{k},i\omega_{m})$, where $\omega_{m}=(2m+1)\pi k_{B}T$
denotes the fermionic Matsubara frequency, is given by Eq. (\ref{eq:gfBCS})
in the main text. For notational convenience and without ambiguity,
hereafter we collectively denote the four-momentum by $k\equiv(\mathbf{k},i\omega_{m})$.
Accordingly, the four-momentum summation is defined as $\sum_{k}\equiv(k_{B}T/V)\sum_{\mathbf{k}}\sum_{i\omega_{m}}.$

By treating the complex fluctuating field $\varphi({\bf x},\tau)$
and the real source field $j_{c}({\bf x},\tau)$ as small expansion
parameters, we expand the trace appearing in the effective action
to obtain,
\begin{equation}
-\text{Tr}\ln\left[-{\cal G}_{j}^{-1}\right]\simeq-\text{Tr}\ln\left[-{\cal G}_{0}^{-1}\right]+\frac{1}{2}\text{Tr}\ln\left\{ \left[\left(\mathcal{G}_{0}\sigma^{-}\right)\varphi^{*}\left({\bf x},\tau\right)+\left(\mathcal{G}_{0}\sigma^{+}\right)\varphi\left({\bf x},\tau\right)+\left(\mathcal{G}_{0}\sigma_{z}\right)j_{c}\left({\bf x},\tau\right)\right]^{2}\right\} ,\label{eq:TraceExpanded}
\end{equation}
and retain all terms up to second order in $\varphi$, $\varphi^{*}$
and $j_{c}$, including their mixed terms. Truncating the expansion
at this Gaussian level and performing a Fourier transformation to
momentum space, we obtain the fluctuation part of the effective action,
\begin{equation}
{\cal S}_{\textrm{eff}}^{(2)}=\frac{1}{2}\sum_{q}\left(\begin{array}{ll}
\varphi_{q}^{*}, & \varphi_{-q}\end{array},j_{c,-q}\right)\left[\begin{array}{ccc}
M_{11}\left(q\right) & M_{12}\left(q\right) & \chi_{1c}\left(q\right)\\
M_{12}\left(q\right) & M_{22}\left(q\right) & \chi_{2c}\left(q\right)\\
\chi_{1c}\left(q\right) & \chi_{2c}\left(q\right) & \chi_{cc}\left(q\right)
\end{array}\right]\left(\begin{array}{c}
\varphi_{q}\\
\varphi_{-q}^{*}\\
j_{c,q}
\end{array}\right),\label{eq:Seff2}
\end{equation}
where $q\equiv(\mathbf{q},i\nu_{n})$ denotes the four-momentum with
the bosonic Matsubara frequency $\nu_{n}\equiv2n\pi k_{B}T$. It is
readily seen that the matrix element $\chi_{cc}(q)=\textrm{Tr}[\mathcal{G}_{0}(k)\sigma_{z}\mathcal{G}_{0}(k+q)\sigma_{z}]$,
obtained by identifying the coefficient of the quadratic term $(1/2)\textrm{Tr}\left[\mathcal{G}_{0}(\mathbf{x},\tau;\mathbf{x}',\tau')\sigma_{z}j_{c}(\mathbf{x}',\tau')\mathcal{G}_{0}(\mathbf{x}',\tau';\mathbf{x},\tau)\sigma_{z}j_{c}(\mathbf{x},\tau)\right]$
after Fourier transformation, corresponds to the bare response to
the source field $j_{c}(\mathbf{x},\tau)$:
\begin{equation}
\chi_{cc}\left(q\right)=\sum_{k}\left[{\cal G}_{11}^{(0)}\left(k\right){\cal G}_{11}^{(0)}\left(k+q\right)-{\cal G}_{12}^{(0)}\left(k\right){\cal G}_{21}^{(0)}\left(k+q\right)-{\cal G}_{21}^{(0)}\left(k\right){\cal G}_{12}^{(0)}\left(k+q\right)+{\cal G}_{22}^{(0)}\left(k\right){\cal G}_{22}^{(0)}\left(k+q\right)\right].
\end{equation}
On the other hand, it is straightforward to identify the four matrix
elements, $M_{ij}(q)$ ($i,j=1,2$) with $M_{22}(q)=M_{11}(-q)$,
as the elements of the inverse $2\times2$ pairing vertex, apart from
an overall minus sign,
\begin{equation}
\boldsymbol{\Gamma}\left(q\right)=-\left[\begin{array}{cc}
M_{11}\left(q\right) & M_{12}\left(q\right)\\
M_{12}\left(q\right) & M_{22}\left(q\right)
\end{array}\right]^{-1}.
\end{equation}
$\mathbf{\boldsymbol{\Gamma}}(q)$ may be interpreted as the unnormalized
Green\textquoteright s function of Cooper pairs within the Gaussian-fluctuation
framework. The corresponding expressions for $M_{ij}(q)$ are well
known in the literature \citep{He2015,Hu2006,Diener2008}: 
\begin{eqnarray}
M_{11}\left(q\right) & = & -\frac{1}{U_{0}}+\sum_{k}{\cal G}_{22}^{(0)}\left(k\right){\cal G}_{11}^{(0)}\left(k+q\right),\\
M_{12}\left(q\right) & = & \sum_{k}{\cal G}_{12}^{(0)}\left(k\right){\cal G}_{12}^{(0)}\left(k+q\right).
\end{eqnarray}
Due to the nonzero mixed susceptibilities 
\begin{equation}
\chi_{1c}\left(q\right)=\textrm{Tr}\left[\mathcal{G}_{0}\left(k\right)\sigma^{-}\mathcal{G}_{0}\left(k+q\right)\sigma_{z}\right]=\sum_{k}\left[{\cal G}_{12}^{(0)}\left(k\right){\cal G}_{11}^{(0)}\left(k+q\right)-{\cal G}_{22}^{(0)}\left(k\right){\cal G}_{12}^{(0)}\left(k+q\right)\right]
\end{equation}
and $\chi_{2c}(q)=\textrm{Tr}[\mathcal{G}_{0}(k)\sigma^{+}\mathcal{G}_{0}(k+q)\sigma_{z}]=\chi_{1c}(-q)$,
the density fluctuation couples to the collective condensate dynamics
encoded in the pairing vertex $\mathbf{\boldsymbol{\Gamma}}(q)$.
This coupling becomes explicit upon integrating out the pair-fluctuation
fields $\varphi$, $\varphi^{*}$ from the fluctuation action in Eq.
(\ref{eq:Seff2}), yielding
\begin{equation}
{\cal \tilde{S}}_{\textrm{eff}}^{(2)}=\frac{1}{2}\sum_{q}j_{c,-q}\left[\chi_{cc}^{\textrm{RPA}}\left(q\right)\right]j_{c,q},
\end{equation}
where
\begin{equation}
\chi_{cc}^{\textrm{RPA}}\left(q\right)=\chi_{cc}\left(q\right)-\left[\begin{array}{cc}
\chi_{1c}\left(q\right), & \chi_{1c}\left(-q\right)\end{array}\right]\left[\begin{array}{cc}
M_{11}\left(q\right) & M_{12}\left(q\right)\\
M_{12}\left(q\right) & M_{22}\left(q\right)
\end{array}\right]^{-1}\left[\begin{array}{c}
\chi_{1c}\left(q\right)\\
\chi_{1c}\left(-q\right)
\end{array}\right]
\end{equation}
\end{widetext}can be identified as the charge-density response with
the inclusion of the dynamical feedback from the condensate. Upon
performing the analytic continuation $i\nu_{n}\rightarrow\omega+i\eta$,
with $\eta=0^{+}$ an infinitesimal positive number, we obtain the
RPA charge-density response function $\chi_{cc}^{\textrm{RPA}}(\mathbf{q},\omega)$,
as given in Eq. (\ref{eq:kappaRPA}) of the main text.

The spin-density response can be derived in an analogous manner. The
only modification is that, in Eq. (\ref{eq:TraceExpanded}), $(\mathcal{G}_{0}\sigma_{z})j_{c}({\bf x},\tau)$
is replaced by $(\mathcal{G}_{0}\sigma_{0})j_{s}({\bf x},\tau)$,
i.e., the Pauli matrix $\sigma_{z}$ is replaced by the identity matrix
$\sigma_{0}$. Hence, the bare spin-density response is given by $\chi_{ss}(q)=\textrm{Tr}[\mathcal{G}_{0}(k)\sigma_{0}\mathcal{G}_{0}(k+q)\sigma_{0}]$,
while the mixed susceptibility becomes $\chi_{1s}(q)=\textrm{Tr}[\mathcal{G}_{0}(k)\sigma^{-}\mathcal{G}_{0}(k+q)\sigma_{0}]=\sum_{k}{\cal G}_{12}^{(0)}(k)[{\cal G}_{11}^{(0)}(k+q)+{\cal G}_{22}^{(0)}(-k-q)]$.
For conventional superconductors with time-reversal symmetry, the
BCS Green's functions satisfy $\mathcal{G}_{11}^{(0)}(k)=-\mathcal{G}_{22}^{(0)}(-k)$,
and consequently the mixed susceptibility in the spin channel, $\chi_{1s}(q)$,
vanishes identically. This relation no longer holds in an altermagnetic
superconductor, where both the altermagnetic coupling and the external
magnetic field break the time-reversal symmetry. As a result, $\chi_{1s}(q)$
becomes nonzero and the spin-density response couples to both the
phonon and Higgs modes.

\begin{figure*}
\begin{centering}
\includegraphics[width=0.5\textwidth]{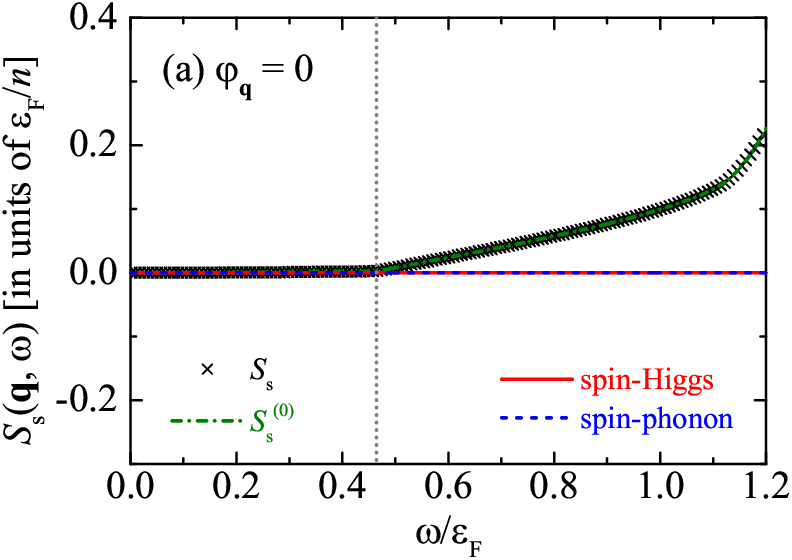}\includegraphics[width=0.5\textwidth]{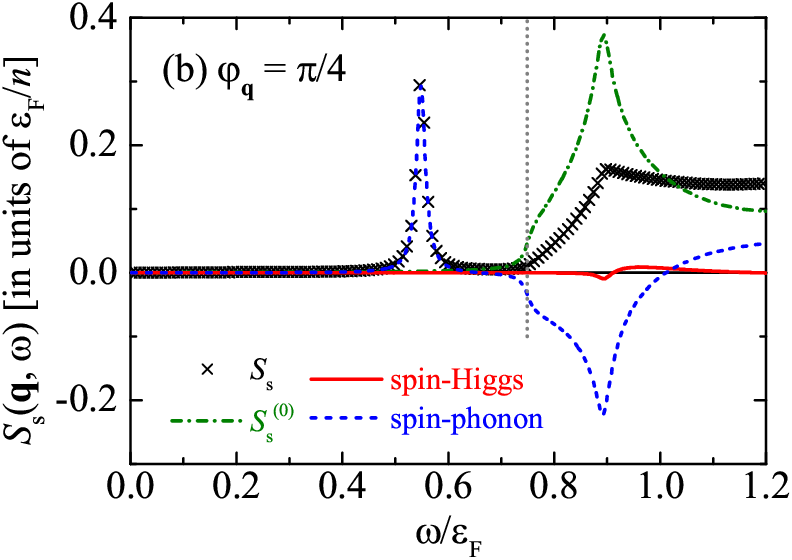}
\par\end{centering}
\caption{\label{figA1: sDSF} Spin DSF $S_{s}(\mathbf{q},\omega)$ along the
two momentum direction, $\varphi_{\mathbf{q}}=0$ (a) and $\varphi_{\mathbf{q}}=+45^{\circ}$
(b), for the fully gapped BCS state with $\lambda=0.7$ and $h=0$.
The total spin-density response $S_{s}(\mathbf{q},\omega)$ (black
crosses) is decomposed into contributions from particle-hole excitations
$S_{s}^{(0)}(\mathbf{q},\omega)$ (green dot-dashed line), the Higgs
mode $S_{\ensuremath{s,\lambda\lambda}}(\mathbf{q},\omega)$ (red
solid line), and the phonon mode $S_{s,\theta\theta}(\mathbf{q},\omega)$
(blue dashed line). The momentum magnitude is fixed at $q=\left|\mathbf{q}\right|=0.5k_{F}$.
The vertical dotted lines in (a) and (b) indicate the pair-breaking
gaps (i.e., the two-particle scattering thresholds), given by $E_{2p}(\varphi_{\mathbf{q}}=0)\simeq2\Delta-\lambda\sqrt{\mu}q-\lambda^{2}\Delta q^{2}/(16\mu)$
and $E_{2p}(\varphi_{\mathbf{q}}=\pi/4)=\Delta\sqrt{4-\lambda^{2}}$,
respectively.}
\end{figure*}

\section{Spin-Higgs and spin-phonon couplings}

To elucidate how the spin response couples to the Higgs and phonon
modes, it is convenient to decompose the pair-fluctuation fields $\varphi_{q}$
and $\varphi_{q}^{*}$ into a real amplitude field $\lambda_{q}$
and a real phase field $\theta_{q}$, up to an unimportant overall
normalization factor: $\varphi_{q}=(\lambda_{q}+i\theta_{q})/\sqrt{2}$
and $\varphi_{q}^{*}=(\lambda_{-q}-i\theta_{-q})/\sqrt{2}$. Under
the presence of the source field $j_{s}(\mathbf{x},\tau)$, the resulting
effective fluctuation action takes the following form,\begin{widetext}
\begin{equation}
{\cal S}_{\textrm{eff}}^{(2)}=\frac{1}{2}\sum_{q}\left(\begin{array}{ll}
\lambda_{-q}, & \theta_{-q}\end{array},j_{s,-q}\right)\left[\begin{array}{ccc}
D_{11}\left(q\right) & iD_{12}\left(q\right) & \chi_{\lambda s}\left(q\right)\\
-iD_{12}\left(q\right) & D_{22}\left(q\right) & i\chi_{\theta s}\left(q\right)\\
\chi_{\lambda s}\left(q\right) & -i\chi_{\theta s}\left(q\right) & \chi_{cc}\left(q\right)
\end{array}\right]\left(\begin{array}{c}
\lambda_{q}\\
\theta_{q}\\
j_{s,q}
\end{array}\right),\label{eq:Seff2spin}
\end{equation}
where $D_{11}=(M_{11}+M_{22})/2+M_{12}$, $D_{12}=(M_{11}-M_{22})/2$,
$D_{22}=(M_{11}+M_{22})/2-M_{12}$, $\chi_{\lambda s}=\chi_{1s}+\chi_{2s}$,
and $\chi_{\theta s}=-\chi_{1s}+\chi_{2s}$. Accordingly, the spin-density
response function within the RPA framework can be expressed as, 
\begin{equation}
\chi_{ss}^{\textrm{RPA}}\left(\mathbf{q},\omega\right)=\chi_{ss}\left(\mathbf{q},\omega\right)+\left[\begin{array}{cc}
\chi_{\lambda s}\left(\mathbf{q},\omega\right), & -i\chi_{\theta s}\left(\mathbf{q},\omega\right)\end{array}\right]\left[\begin{array}{cc}
\Gamma_{\lambda\lambda}\left(\mathbf{q},\omega\right) & \Gamma_{\lambda\theta}\left(\mathbf{q},\omega\right)\\
\Gamma_{\theta\lambda}\left(\mathbf{q},\omega\right) & \Gamma_{\theta\theta}\left(\mathbf{q},\omega\right)
\end{array}\right]\left[\begin{array}{c}
\chi_{\lambda s}\left(\mathbf{q},\omega\right)\\
i\chi_{\theta s}\left(\mathbf{q},\omega\right)
\end{array}\right].
\end{equation}
In the amplitude--phase representation, the pairing vertex correspondingly
takes the form,
\begin{equation}
\left[\begin{array}{cc}
\Gamma_{\lambda\lambda} & \Gamma_{\lambda\theta}\\
\Gamma_{\theta\lambda} & \Gamma_{\theta\theta}
\end{array}\right]=-\left[\begin{array}{cc}
D_{11} & iD_{12}\\
-iD_{12} & D_{22}
\end{array}\right]^{-1}=\left[\begin{array}{cc}
\left(\Gamma_{11}+\Gamma_{22}\right)/2+\Gamma_{12} & i\left(\Gamma_{11}-\Gamma_{22}\right)/2\\
-i\left(\Gamma_{11}-\Gamma_{22}\right)/2 & \left(\Gamma_{11}+\Gamma_{22}\right)/2-\Gamma_{12}
\end{array}\right].
\end{equation}
\end{widetext}It is then useful to decompose the spin dynamical structure
factor (DSF) into distinct contributions, 
\begin{equation}
S_{s}=S_{s}^{(0)}+S_{s,\lambda\lambda}+2S_{s,\lambda\theta}+S_{s,\theta\theta},
\end{equation}
where 
\begin{eqnarray}
S_{s}^{(0)} & = & -\frac{1}{\pi}\left[1+n_{B}\left(\omega\right)\right]\textrm{Im}\chi_{ss},\\
S_{s,\lambda\lambda} & = & -\frac{1}{\pi}\left[1+n_{B}\left(\omega\right)\right]\textrm{Im}\left[\chi_{\lambda s}\Gamma_{\lambda\lambda}\chi_{\lambda s}\right],\\
S_{s,\lambda\theta} & = & -\frac{1}{\pi}\left[1+n_{B}\left(\omega\right)\right]\textrm{Im}\left[\chi_{\lambda s}\Gamma_{\lambda\theta}\chi_{\theta s}\right],\\
S_{s,\theta\theta} & = & -\frac{1}{\pi}\left[1+n_{B}\left(\omega\right)\right]\textrm{Im}\left[\chi_{\theta s}\Gamma_{\theta\theta}\chi_{\theta s}\right].
\end{eqnarray}
These four terms correspond, respectively, to contributions from particle-hole
excitations, the Higgs mode, Higgs-phonon mode coupling, and the phonon
mode. In general, we find that the contribution arising from the Higgs-phonon
mode coupling is subleading and can therefore be neglected to a good
approximation, such that $S_{s,\lambda\theta}\simeq0$.

In Fig. \ref{figA1: sDSF}, we show the different contributions to
the spin DSF for a fully gapped BCS state with $\lambda=0.7$ and
$h=0$, at $\left|\mathbf{q}\right|=0.5k_{F}$ and for two momentum
directions, $\varphi_{\mathbf{q}}=0$ and $\varphi_{\mathbf{q}}=+45^{\circ}$.
When the momentum is directed along the $x$-axis (or $y$-axis),
no spin response is observed within the pair-breaking gap, i.e., for
$\omega<0.5\varepsilon_{F}$, owing to the accidental vanishing of
the coupling $\chi_{1s}=0$ along this highly symmetric direction.
In contrast, for momentum directed along the diagonal, $\varphi_{\mathbf{q}}=+45^{\circ}$,
a finite spin response emerges within the pair-breaking gap, $\omega<0.7\varepsilon_{F}$,
which originates entirely from the phonon mode. Above the gap, the
bare spin response associated with particle-hole excitations becomes
the dominant contribution. Notably, this contribution completely compensates
the anti-peak arising from the phonon mode around $\omega\sim0.9\varepsilon_{F}$,
resulting in a broader shoulder-like feature in this frequency range.

\begin{figure}
\begin{centering}
\includegraphics[width=0.5\textwidth]{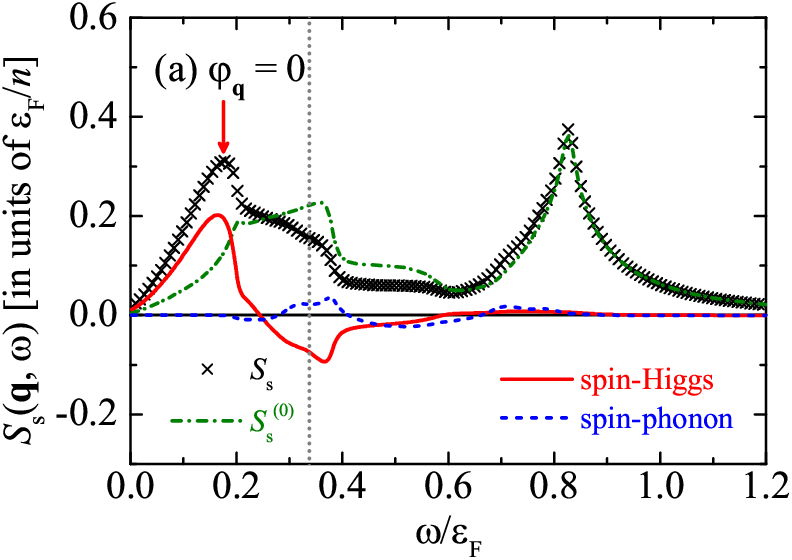}
\par\end{centering}
\begin{centering}
\includegraphics[width=0.5\textwidth]{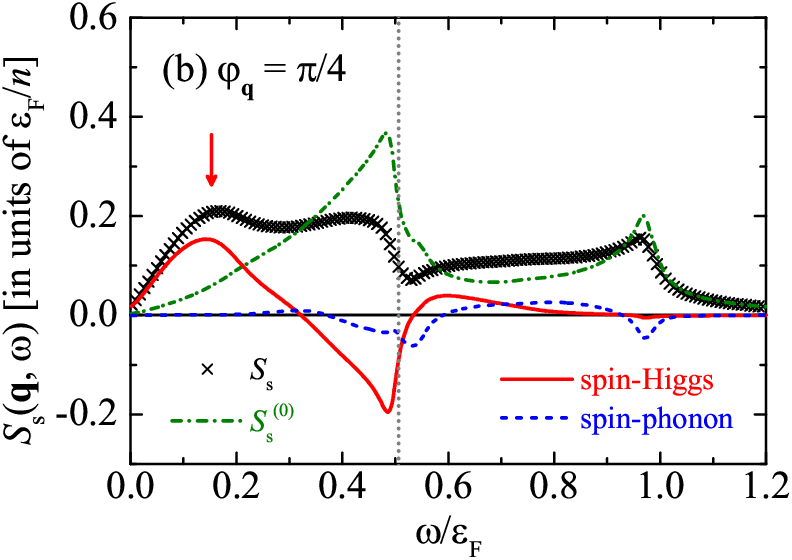}
\par\end{centering}
\caption{\label{figA2: sDSF} Spin DSF $S_{s}(\mathbf{q},\omega)$ along the
two momentum direction, $\varphi_{\mathbf{q}}=0$ (a) and $\varphi_{\mathbf{q}}=+45^{\circ}$
(b), for the polarized BCS state with $\lambda=0.7$ and $h=0$.15$\varepsilon_{F}$.
The total spin-density response $S_{s}(\mathbf{q},\omega)$ (black
crosses) is decomposed into contributions from particle-hole excitations
$S_{s}^{(0)}(\mathbf{q},\omega)$ (green dot-dashed line), the Higgs
mode $S_{\ensuremath{s,\lambda\lambda}}(\mathbf{q},\omega)$ (red
solid line), and the phonon mode $S_{s,\theta\theta}(\mathbf{q},\omega)$
(blue dashed line). The red arrows indicate the position of the Higgs-mode
peak within the pair-breaking gap. The momentum magnitude is fixed
at $q=\left|\mathbf{q}\right|=0.3k_{F}$. The vertical dotted lines
in (a) and (b) indicate the pair-breaking gaps (i.e., the two-particle
scattering thresholds), given by $E_{2p}(\varphi_{\mathbf{q}}=0)\simeq2\Delta-\lambda\sqrt{\mu}q-\lambda^{2}\Delta q^{2}/(16\mu)$
and $E_{2p}(\varphi_{\mathbf{q}}=\pi/4)=\Delta\sqrt{4-\lambda^{2}}$,
respectively.}
\end{figure}

Therefore, for a fully gapped BCS state, we find that the coupling
between the spin response and the collective modes is dominated by
the phonon contribution. This behavior, however, changes upon entering
the gapless polarized BCS phase with Bogoliubov Fermi surfaces (BFSs).
In this regime, the phonon mode can be suppressed or even destroyed
by gapless particle-hole excitations around the BFSs for certain momentum
directions. Consequently, we expect the coupling between the spin
response and the collective modes to be dominated instead by the in-gap
feature of the Higgs mode.

To verify this expectation, we present in Fig. \ref{figA2: sDSF}
the different contributions to the spin DSF for a polarized BCS state
with $\lambda=0.7$ and $h=0.15\varepsilon_{F}$. We consider the
same two momentum directions, $\varphi_{\mathbf{q}}=0$ and $\varphi_{\mathbf{q}}=+45^{\circ}$,
with $\left|\mathbf{q}\right|=0.3k_{F}$. For both momentum directions,
the contribution of the phonon mode to the spin DSF (blue dashed lines)
is found to be subleading. At low frequencies, the spin DSF is dominated
by the Higgs mode and exhibits a pronounced low-energy peak at $\omega\sim0.15\varepsilon_{F}$.
This peak coincides with the in-gap peak observed in the Higgs spectral
function, as indicated by the red arrows in the figure. 

At frequencies above this low-energy peak, the contribution from particle-hole
excitations becomes dominant. Interestingly, in the same frequency
regime, the Higgs mode develops a negative contribution to the spin
DSF, indicating out-of-phase dynamics between the particle-hole excitations
and the Higgs mode. This behavior can be understood from the dynamical
feedback of the condensate's collective fluctuations, which generally
acts in opposition to the bare response of the particle-hole excitations.

In summary, in the fully gapped BCS state, the coupling between the
spin response and the collective modes is mediated predominantly by
the phonon mode. In contrast, in the gapless polarized BCS phase where
the phonon mode becomes less well-defined, the Higgs mode plays a
dominant role in mediating the in-gap coupling to the spin response.

\begin{figure}[b]
\begin{centering}
\includegraphics[width=0.5\textwidth]{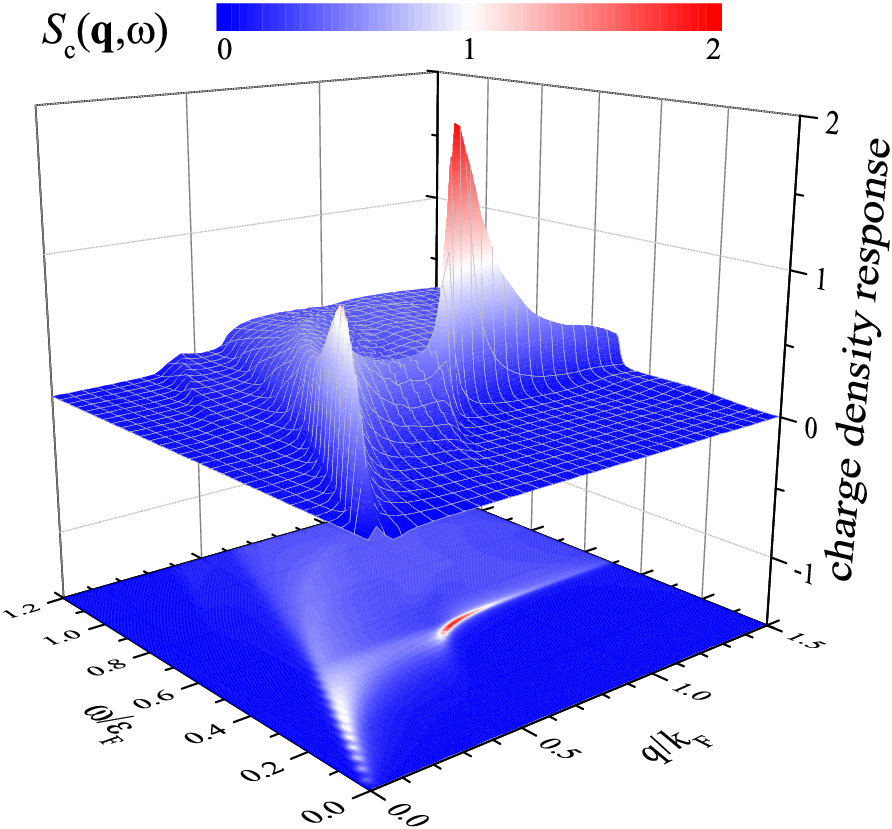}
\par\end{centering}
\caption{\label{figA3: cDSFp025pi} Two-dimensional contour plot of the charge
DSF $S_{c}(\mathbf{q},\omega)$ along the diagonal momentum direction,
$\varphi_{\mathbf{q}}=+45^{\circ}$, for the polarized BCS state with
$\lambda=0.7$ and $h=0.15\varepsilon_{F}$.}
\end{figure}

\section{Undamped phonon mode at $\varphi_{\mathbf{q}}=+45^{\circ}$}

In the polarized BCS state, we find an undamped phonon mode along
the diagonal momentum direction near $q=\mathbf{\left|q\right|}\simeq0.8k_{F}$,
arising from the restricted phase space available for particle-hole
excitations around the BFSs. This feature is particularly evident
in the charge DSF, as shown in Fig. \ref{figA3: cDSFp025pi}, where
sharp peaks emerge at $\mathbf{\left|q\right|}\simeq0.8k_{F}$.
\end{document}